\documentclass[aps,prb,twocolumn,showpacs,amsmath,amssymb,superscriptaddress]{revtex4-2}

\usepackage{tabularx}
\usepackage{bm}
\usepackage{graphicx}
\usepackage{tikz}

\usepackage{hyperref}
\hypersetup{colorlinks=true,urlcolor= blue,citecolor=blue,linkcolor= blue,bookmarks=true,bookmarksopen=false}

\usepackage{color}

\usepackage{amsmath,mathtools}
\usepackage{multirow}
\usepackage{dcolumn}
\usepackage{amssymb,amscd,xypic,bm,wasysym}
\usepackage{float}
\usepackage{cleveref}
\usepackage[caption=false,position=top,captionskip=0pt,farskip=0pt]{subfig}
\usepackage{soul}

\let\oldhat\hat
\renewcommand{\hat}[1]{\oldhat{\mathbf{#1}}}
\renewcommand{\vec}[1]{\mathbf{#1}}

\newcommand{\ham}{\mathcal{H}}
\newcommand{\cc}{c^{\dagger}}
\newcommand{\de}{\Delta}

\begin{document}

\title{Superconducting triangular islands as a platform for manipulating Majorana zero modes}

\author{Aidan Winblad}
\affiliation{Department of Physics, Colorado State University, Fort Collins, CO 80523, USA}

\author{Hua Chen}
\affiliation{Department of Physics, Colorado State University, Fort Collins, CO 80523, USA}
\affiliation{School of Advanced Materials Discovery, Colorado State University, Fort Collins, CO 80523, USA}

\begin{abstract}
Current proposals for topological quantum computation (TQC) based on Majorana zero modes (MZM) have mostly been focused on coupled-wire architecture which can be challenging to implement experimentally. To explore alternative building blocks of TQC, in this work we study the possibility of obtaining robust MZM at the corners of triangular superconducting islands, which often appear spontaneously in epitaxial growth. We first show that a minimal three-site triangle model of spinless $p$-wave superconductor allows MZM to appear at different pairs of vertices controlled by a staggered vector potential, which may be realized using coupled quantum dots and can already demonstrate braiding. For systems with less fine-tuned parameters, we suggest an alternative structure of a ``hollow" triangle subject to uniform supercurrents or vector potentials, in which MZM generally appear when two of the edges are in a different topological phase from the third. We also discuss the feasibility of constructing the triangles using existing candidate MZM systems and of braiding more MZM in networks of such triangles.
\end{abstract}

\maketitle

\emph{Introduction.---}For more than twenty years, Majorana zero modes (MZM) in condensed matter systems have been highly sought after due to their potential for serving as building blocks of topological quantum computation, thanks to their inherent robustness against decoherence and non-Abelian exchange statistics \cite{ivanovNonAbelianStatisticsHalfQuantum2001, kitaevFaulttolerantQuantumComputation2003, nayakNonAbelianAnyonsTopological2008, aliceaNonAbelianStatisticsTopological2011, aasenMilestonesMajoranaBasedQuantum2016}. MZM were originally proposed to be found in half-quantum vortices of two-dimensional (2D) topological \textit{p}-wave superconductors and at the ends of 1D spinless \textit{p}-wave superconductors \cite{readPairedStatesFermions2000, kitaevUnpairedMajoranaFermions2001}. Whether a pristine \textit{p}-wave superconductor \cite{brisonPWaveSuperconductivityDVector2021} has been found is still under debate. However, innovative heterostructures proximate to ordinary $s$-wave superconductors have been proposed to behave as effective topological superconductors in both 1D and 2D. These include, for example, semiconductor nanowires subject to magnetic fields \cite{mourikSignaturesMajoranaFermions2012, rokhinsonFractionalJosephsonEffect2012, dengAnomalousZeroBiasConductance2012}, ferromagnetic atomic spin chains \cite{choyMajoranaFermionsEmerging2011, brauneckerInterplayClassicalMagnetic2013, klinovajaTopologicalSuperconductivityMajorana2013,nadj-pergeProposalRealizingMajorana2013,nadj-pergeObservationMajoranaFermions2014,schneiderPrecursorsMajoranaModes2022}, 3D topological insulators \cite{fuSuperconductingProximityEffect2008, hosurMajoranaModesEnds2011, potterEngineeringMathitipSuperconductor2011, veldhorstMagnetotransportInducedSuperconductivity2013}, quantum anomalous Hall insulators \cite{chenQuasionedimensionalQuantumAnomalous2018, zengQuantumAnomalousHall2018, xieCreatingLocalizedMajorana2021}, quasi-2D spin-orbit-coupled superconductors with a perpendicular Zeeman field \cite{oregHelicalLiquidsMajorana2010, sauGenericNewPlatform2010, lutchynSearchMajoranaFermions2011, potterTopologicalSuperconductivityMajorana2012, liTwodimensionalChiralTopological2016, leiUltrathinFilmsSuperconducting2018}, and planar Josephson junctions \cite{black-schafferMajoranaFermionsSpinorbitcoupled2011, pientkaSignaturesTopologicalPhase2013, hellTwoDimensionalPlatformNetworks2017, fornieriEvidenceTopologicalSuperconductivity2019, renTopologicalSuperconductivityPhasecontrolled2019, scharfTuningTopologicalSuperconductivity2019, zhouPhaseControlMajorana2020}, etc. It has been a challenging task to decisively confirm the existence of MZM in the various experimental systems due to other competing mechanisms that can potentially result in similar features as MZM do in different probes \cite{xuExperimentalDetectionMajorana2015, albrechtExponentialProtectionZero2016, sunMajoranaZeroMode2016, wangEvidenceMajoranaBound2018, jackObservationMajoranaZero2019, fornieriEvidenceTopologicalSuperconductivity2019, renTopologicalSuperconductivityPhasecontrolled2019, mannaSignaturePairMajorana2020}. Other proposals for constructing Kitaev chains through a bottom-up approach, based on, e.g. magnetic tunnel junctions proximate to spin-orbit-coupled superconductors \cite{fatinWirelessMajoranaBound2016}, and quantum dots coupled through superconducting links \cite{sauRealizingRobustPractical2012,leijnseParityQubitsPoor2012,dvirRealizationMinimalKitaev2023} are therefore promising. In particular, the recent experiment \cite{dvirRealizationMinimalKitaev2023} of a designer minimal Kitaev chain based on two quantum dots coupled through tunable crossed Andreev reflections (CAR) offers a compelling route towards MZM platforms based on exactly solvable building blocks.

In parallel with the above efforts of realizing MZM in different materials systems, scalable architectures for quantum logic circuits based on MZM have also been intensely studied over the past decades. A major proposal among these studies is to build networks of T-junctions, which are minimal units for swapping a pair of MZM hosted at different ends of a junction, that allow braiding-based TQC \cite{aasenMilestonesMajoranaBasedQuantum2016}. Alternatively, networks based on coupled wires forming the so-called tetrons and hexons, aiming at measurement-based logic gate operations  \cite{karzigScalableDesignsQuasiparticlepoisoningprotected2017}, have also been extensively investigated. To counter the technical challenges of engineering networks with physical wires or atomic chains, various ideas based on effective Kitaev chains, such as quasi-1D systems in thin films \cite{potterMultichannelGeneralizationKitaev2010}, cross Josephson junctions \cite{zhouPhaseControlMajorana2020}, scissor cuts on a quantum anomalous Hall insulator \cite{xieCreatingLocalizedMajorana2021}, and rings of magnetic atoms \cite{liManipulatingMajoranaZero2016}, etc. have been proposed. However, due to the same difficulty of obtaining or identifying genuine MZM in quasi-1D systems mentioned above, it remains unclear how practical these strategies are in the near future. These challenges, along with the advancements in building designer minimal Kitaev chains, motivate us to explore new MZM platforms that are not based on bulk-boundary correspondence: In small systems with only a few fermion degrees of freedom, discussing the emergence of MZM due to bulk-boundary correspondence is less meaningful. Instead, it is easier to fine-tune system parameters based on exactly solvable models to realize well-behaved MZM.

Additionally, in this Letter we highlight triangular superconducting islands as a promising structural unit for manipulating MZM. Unique geometries combined with simple protocols of control parameters can greatly facilitate MZM creation and operations \cite{liManipulatingMajoranaZero2016, pahomiBraidingMajoranaCorner2020, zhangsb_2020_1, zhangsb_2020_2}. We also note that triangles naturally break 2D inversion symmetry and do not present a straightforward strategy for morphing into either 1D or 2D structures with periodic boundary conditions, implying different bulk-boundary physics from other quasi-2D structures. Finally, it is worth mentioning that triangular islands routinely appear spontaneously in epitaxial growth \cite{pietzschSpinResolvedElectronicStructure2006} on close-packed atomic surfaces.

\begin{figure}[ht]
  \hspace{-18pt}
  \subfloat[]{\includegraphics[width=2.0 in]{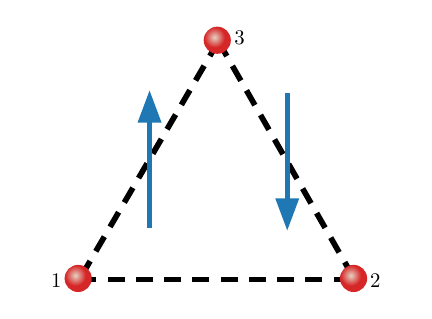}}
  \subfloat[]{\includegraphics[width=1.6 in]{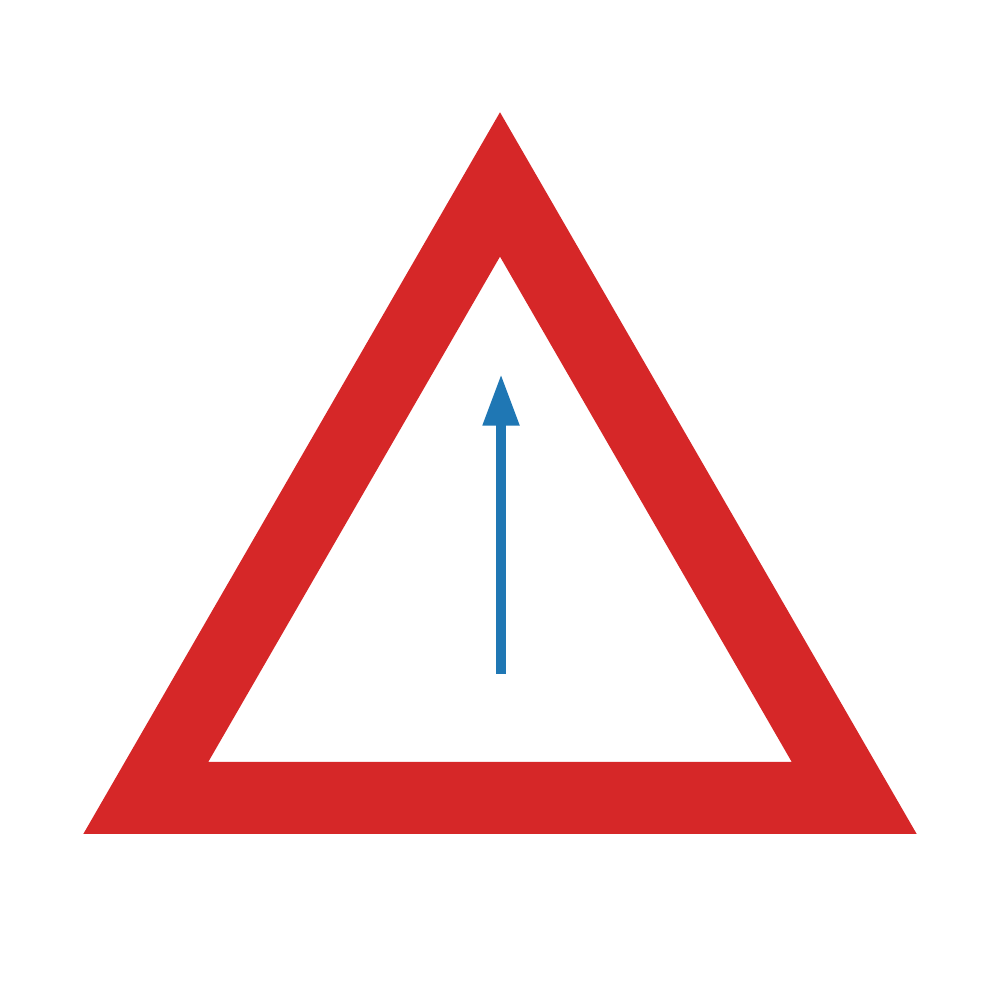}}
  \caption{Schematics of two triangle structures proposed in this work. (a) Three-site Kitaev triangle with bond-dependent Peierls phases. (b) Hollow triangular island with a uniform vector potential.}
  \label{fig:triangles}
\end{figure}

In this Letter we propose two triangular geometry designs that are pertinent to different experimental platforms. The first is an exactly solvable ``Kitaev triangle'' model consisting of three fermion sites. The Kitaev triangle hosts MZM at different pairs of vertices controlled by Peierls phases on the three edges [Fig.~\ref{fig:triangles} (a)], that is not due to topological bulk-boundary correspondence, and can realize the braiding of two MZM. The second is finite-size triangles with a hollow interior [Fig.~\ref{fig:triangles} (b)] under a uniform vector potential, which tunes its individual edges into different topological phases. Compared to existing proposals based on vector potentials or supercurrents \cite{romitoManipulatingMajoranaFermions2012,takasanSupercurrentinducedTopologicalPhase2022,Hyart_2013,Dmytruk_2019}, our design explores the utility of geometry rather than the individual control of superconducting nanowires. We also discuss scaled-up networks of triangles for implementing braiding operations of MZM.

\begin{figure}[ht]
  \subfloat[]{\includegraphics[width=0.3\textwidth]{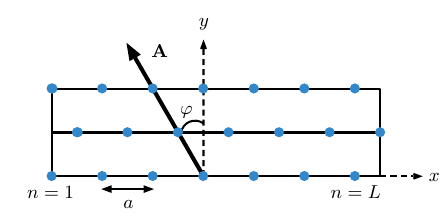}}\\
  \hspace{10pt}
  \subfloat[]{\includegraphics[width=0.355\textwidth]{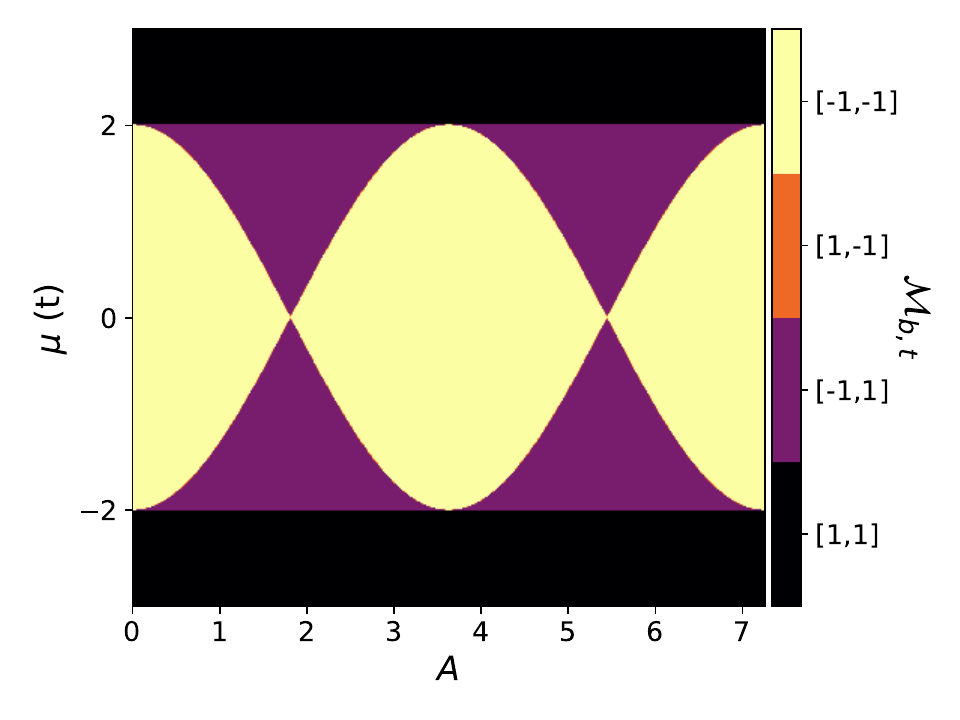}} \\
  \hspace{-20pt}
  \subfloat[]{\includegraphics[width=0.30\textwidth]{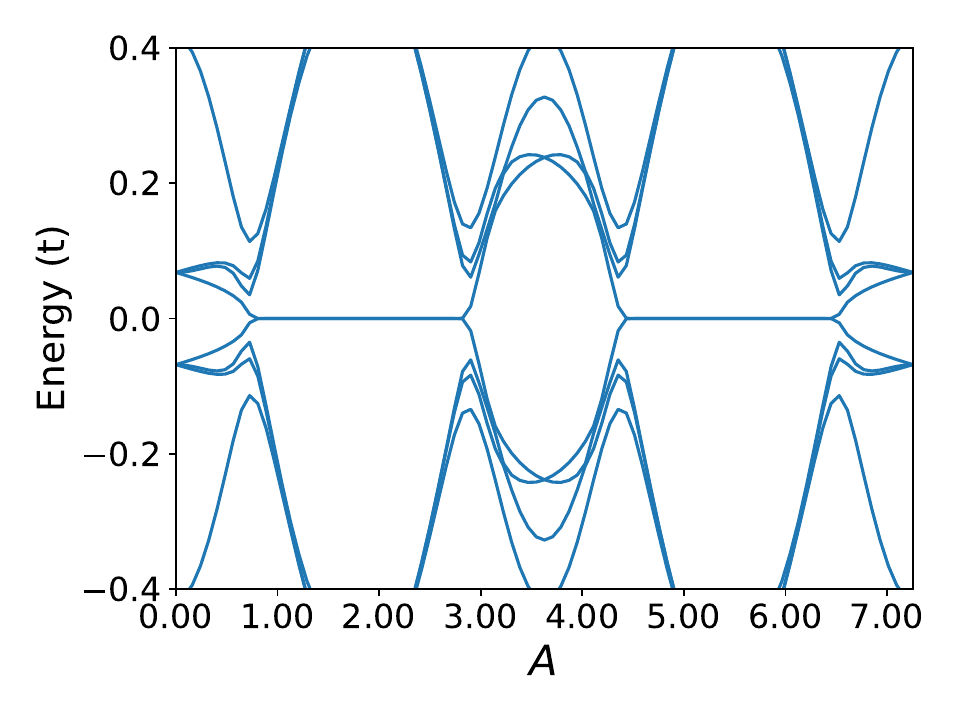}}
  \caption{(a) Schematic illustration of a finite-width ($W=3$ here) ribbon based on the triangular lattice in the presence of a vector potential $\mathbf A = A(-\sin\varphi \hat{x} + \cos\varphi \hat{y})$. (b) Topological phase diagram for a $W=1$ triangular chain obtained by superimposing the $\mathcal{M}_{b,t}(A, \mu)$ ($b$-bottom edge, $t-$top edges) plots of 1D chains with $\mathbf A = A\hat{y}$ (bottom edge) and $\mathbf A = A(\frac{\sqrt{3}}{2}\hat{x}+\frac{1}{2}\hat{y})$ (top edges). Color scheme: black---$[\mathcal{M}_b,\mathcal{M}_t]=[1,1]$, yellow---$[-1,-1]$, purple---$[-1,1]$, orange---$[1,-1]$ (not present in this case) (c) Near-gap BdG eigen-energies vs $A$ for a finite triangle with edge length $L = 50$, $W=1$, and $\mu=1.6$. $t=\Delta =1$ in all calculations.}
  \label{fig: pd}
\end{figure}

\emph{Kitaev triangle.---}In this section we present an exactly solvable minimal model with three sites forming a ``Kitaev triangle" that can host MZM at different pairs of vertices controlled by Peierls phases on the edges. The Bogoliubov-de Gennes (BdG) Hamiltonian includes complex hopping and $p$-wave pairing between three spinless fermions forming an equilateral triangle [Fig.~\ref{fig:triangles} (a)]:
\begin{equation}\label{eq:HBdG}
  \ham = \sum_{\langle j l \rangle} (-te^{i\phi_{jl}}\cc_{j} c_l + \de e^{i\theta_{jl}} c_{j} c_l + {\rm h.c.}) - \sum_{j} \mu \cc_j c_j,
\end{equation}
where $t$ is the hopping amplitude, $\de$ is the amplitude of the (2D) $p$-wave pairing, $\mu$ is the chemical potential, $\theta_{jl}$ is the azimuthal angle of $\mathbf r_{jl} = \mathbf r_l - \mathbf r_j$ (the $x$ axis is chosen to be along $\mathbf r_{12}$), consistent with $\{c^\dag_l, c^\dag_j\} = 0$. $\phi_{jl}$ is the Peierls phase due to a bond-dependent vector potential $\mathbf A$ to be specified below (the nearest neighbor distance $a$ is chosen to be the length unit and $e=\hbar=1$ hereinbelow): $\phi_{jl} = \int_{\mathbf r_j}^{\mathbf r_{l}} \vec{A} \cdot d\vec{l} = -\phi_{lj}$. We have chosen a gauge so that the vector potential only appears in the normal part of the Hamiltonian \cite{DeGennes_book}, and the $p$-wave gap $\Delta$ is assumed to be an effective one induced by proximity to a neighboring superconductor, on which the vector potential has negligible influence. The minimal model may be realized as an effective low-energy model of carefully engineered mesoscopic superconductor devices, such as that made by quantum dots connected by superconducting islands \cite{dvirRealizationMinimalKitaev2023}. Rewriting $\mathcal{H}$ in the Majorana fermion basis $a_{j} = c_j + c^\dag_j$, $b_j = \frac{1}{i}(c_j - c^\dag_j)$ and specializing to the Kitaev limit $t=\de$, $\mu=0$, we can obtain explicit conditions for getting MZM at different sites \cite{supp}. For example, first let $\phi_{12} = 0$ so that sites 1 and 2 alone form a minimal Kitaev chain with $\mathcal{H}_{12} = itb_1a_2$ and hosting MZM $a_1$ and $b_2$. Then one can set $\phi_{23}$ and $\phi_{31}$ so that all terms involving the above two Majorana operators cancel out. Solving the corresponding equations gives $\phi_{23} = -\pi/3$ and $\phi_{31} =-\phi_{13} = -\pi/3$. The three Peierls phases can be realized by the following staggered vector potential
\begin{equation}\label{eq:Astep}
  \vec{A} =\left[1-2\Theta(x)\right]\frac{2 \pi}{3\sqrt{3}} \hat{y}
\end{equation}
where $\Theta(x)$ is the Heavisde step function. The above condition for MZM localized at triangle corners can be generalized to Kitaev chains forming a triangular loop, as well as to finite-size triangles of 2D spinless $p$-wave superconductors in the Kitaev limit, as the existence of $a_1$ and $b_2$ are only dictated by the vector potential near the corresponding corners. It should be noted that in the latter case, 1D edge states will arise when the triangle becomes larger, and effectively diminish the gap that protects the corner MZM. In this sense, the gap that protects the MZM in the Kitaev triangle model, defined by the energies of the first excited states $\pm (1-\frac{\sqrt{2}}{2})t\approx \pm 0.29 t$ \cite{supp}, is due to finite size effects. On the other hand, for the longer Kitaev chain, another pair of MZM will appear near the two bottom vertices which can be understood using a topological argument given in the next section. In this sense, the MZM in the Kitaev triangle here are not due to topological bulk-boundary correspondence [the point of $A = \frac{2\pi}{3\sqrt{3}}$ and $\mu=0$ sits in the trivial phase in Fig.~\ref{fig: pd} (a)].

We next show that the minimal Kitaev triangle suffices to demonstrate braiding of MZM. To this end we consider a closed parameter path linearly interpolating between the following sets of values of $\phi_{jl}$:
\begin{eqnarray}
  (\phi_{12},\phi_{23},\phi_{31}):~\bm \phi_1\rightarrow \bm \phi_2 \rightarrow \bm \phi_3 \rightarrow \bm \phi_1
\end{eqnarray}
with $\bm \phi_1 = \left(0,-\frac{\pi}{3},-\frac{\pi}{3}\right )$, $\bm \phi_2 = \left(-\frac{\pi}{3},-\frac{\pi}{3}, 0\right)$, $\bm \phi_3 = \left(-\frac{\pi}{3}, 0, -\frac{\pi}{3}\right )$. It is straightforward to show that at $\bm \phi_{2}$ and $\bm \phi_3$ there are MZM located at sites $1,3$ and $2,3$, respectively. Therefore the two original MZM at sites $1,2$ should switch their positions at the end of the adiabatic evolution.

\begin{figure}[ht]
	\centering
  \hspace{-18pt}
  \subfloat[]{\includegraphics[width=2.2 in]{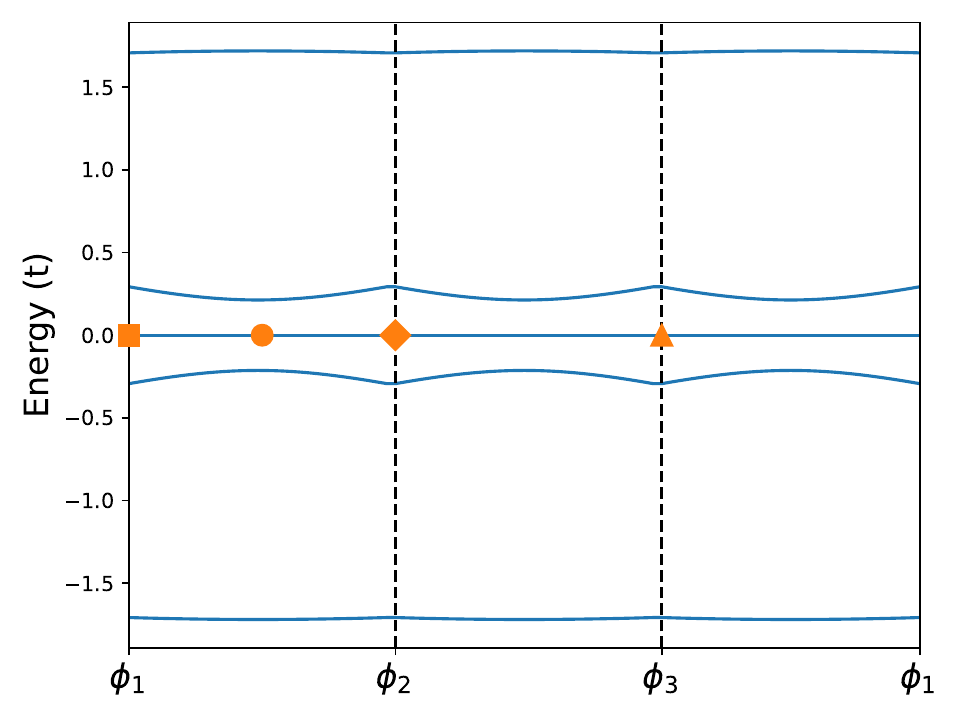}}\\
  \subfloat[]{\includegraphics[width=2.8 in]{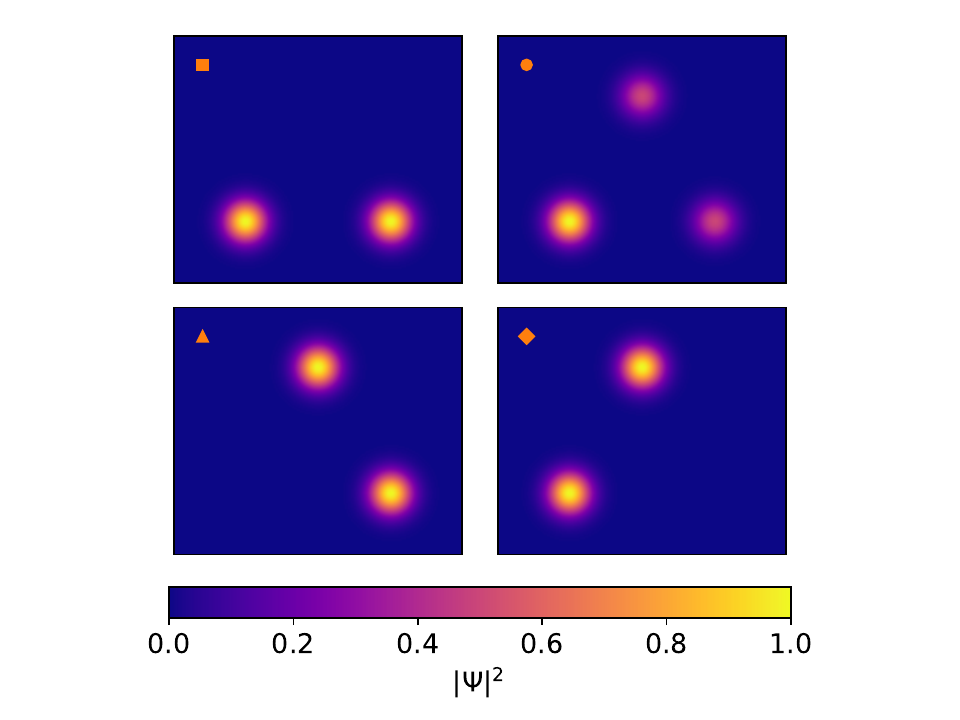}}
	\caption{(a) Evolution of the eigenvalues of the 3-site Kitaev triangle along the closed parameter path for $\phi$ on the three edges. (b) MZM wavefunctions at different points of the parameter path. Clockwise from the upper left panel: $\bm \phi_1 \rightarrow \frac{1}{2}(\bm \phi_1 + \bm \phi_2)\rightarrow \bm \phi_2\rightarrow \bm \phi_3$.}
	\label{fig:3eig}
\end{figure}

Fig.~\ref{fig:3eig} shows that the MZM stays at zero energy throughout the parameter path that interchanges their positions. In \cite{supp} we proved the exact degeneracy of the MZM along the path \cite{FU_2021}. To show that such an operation indeed realizes braiding, we explicitly calculated the many-body Berry phase of the evolution \cite{supp,aliceaNonAbelianStatisticsTopological2011,liManipulatingMajoranaZero2016} and found the two degenerate many-body ground states acquire a $\frac{\pi}{2}$ difference in their Berry phases as expected \cite{aliceaNonAbelianStatisticsTopological2011}. Compared to the minimum T-junction model with four sites \cite{aliceaNonAbelianStatisticsTopological2011,Pandey_2023}, our Kitaev triangle model only requires three sites to achieve braiding between two MZM, and is potentially easier to engineer experimentally.

\emph{Hollow triangles.---}For systems with less fine-tuned Hamiltonians than the minimal model in the previous section, it is more instructive to search for MZM based on topological bulk-boundary correspondence. In this section we show that MZM generally appear at the corners of a hollow triangle, which can be approximated by joining three finite-width chains or ribbons whose bulk topology is individually tuned by the same uniform vector potential.

To this end, we first show that topological phase transitions can be induced by a vector potential in a spinless $p$-wave superconductor ribbon as illustrated in Fig.~\ref{fig: pd} (a). In comparison with similar previous proposals that mostly focused on vector potentials or supercurrents flowing along the chain \cite{romitoManipulatingMajoranaFermions2012, takasanSupercurrentinducedTopologicalPhase2022}, we consider in particular the tunability by varying the direction of the vector potential relative to the length direction of the ribbon, which will become instrumental in a triangular structure.

Consider Eq.~\eqref{eq:HBdG} on a triangular lattice defined by unit-length lattice vectors $(\mathbf a_1, \mathbf a_2) = (\hat{x}, \frac{1}{2}\hat{x} + \frac{\sqrt{3}}{2}\hat{y})$ with $W$ unit cells along $\mathbf a_2$ but infinite unit cells along $\mathbf a_1$, and assume the Peierls phases are due to a uniform vector potential $\mathbf A$ so that $\phi_{jl} = \mathbf A\cdot \mathbf r_{jl}$. The Hamiltonian is periodic along $x$ and can be Fourier transformed through $\cc_{m,n} = \dfrac{1}{\sqrt{N}} \sum_{k} \cc_{k,n} e^{-i km}$, where $m,n$ label the lattice sites as $\mathbf r_{m,n} = m\mathbf a_1 + n \mathbf a_2$. The resulting momentum space Hamiltonian \cite{supp} can then be used to calculate the Majorana number \cite{kitaevUnpairedMajoranaFermions2001,liTopologicalSuperconductivityInduced2014} $\mathcal{M}$ of the 1D ribbon. When $\mathcal{M} = -1$, the 1D system is in a nontrivial topological phase with MZM appearing at open ends of semi-infinite ribbons, and otherwise for $\mathcal{M} = 1$.

\begin{figure*}[ht]
  \subfloat[]{\includegraphics[height=135pt]{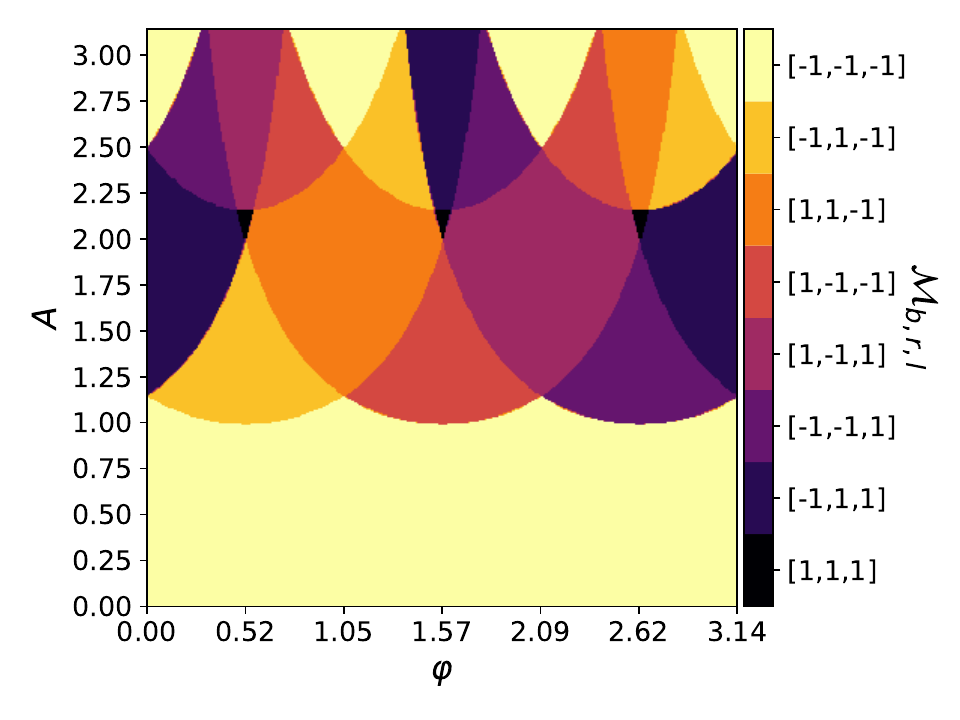}}
  \subfloat[]{\includegraphics[height=135pt]{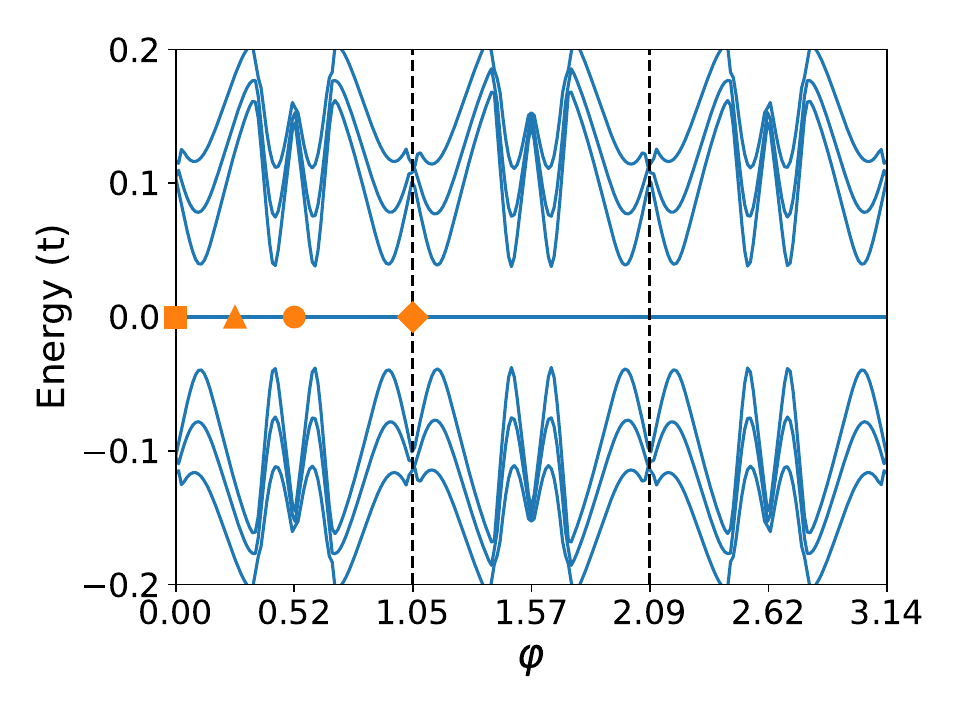}} \\
  \subfloat[]{\includegraphics[height=100pt]{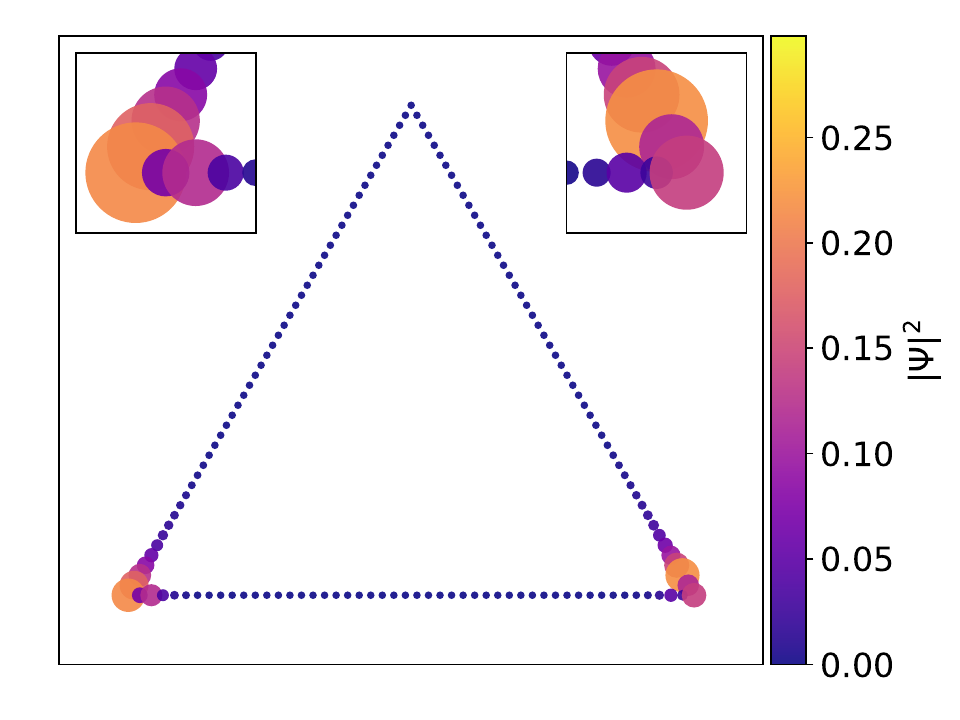}}\hspace{-30.0pt}
  \subfloat[]{\includegraphics[height=100pt]{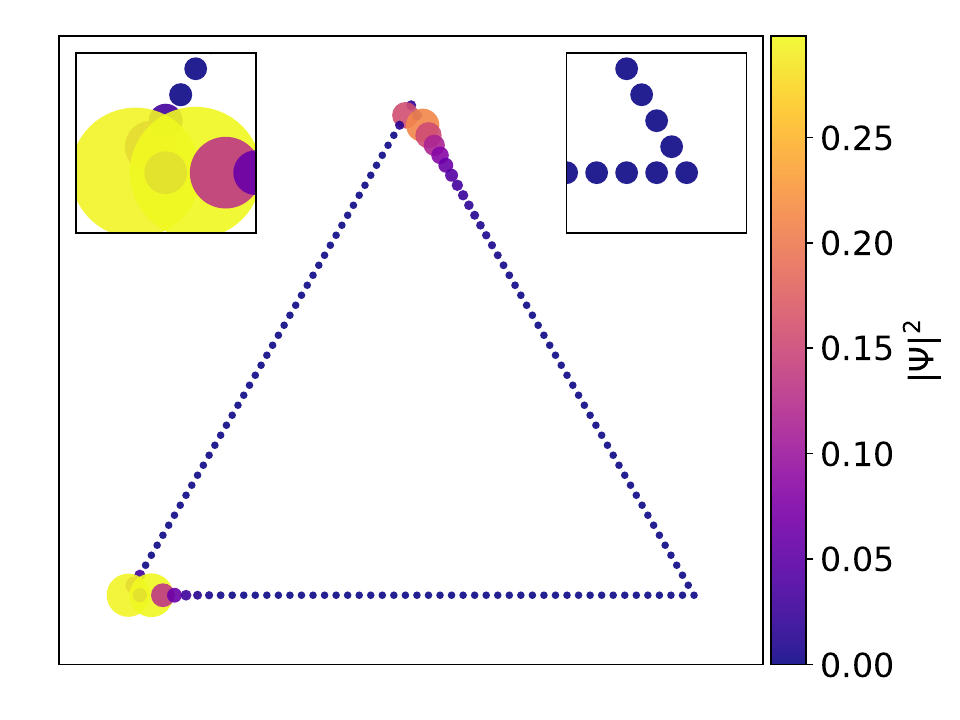}}\hspace{-30.0pt}
  \subfloat[]{\includegraphics[height=100pt]{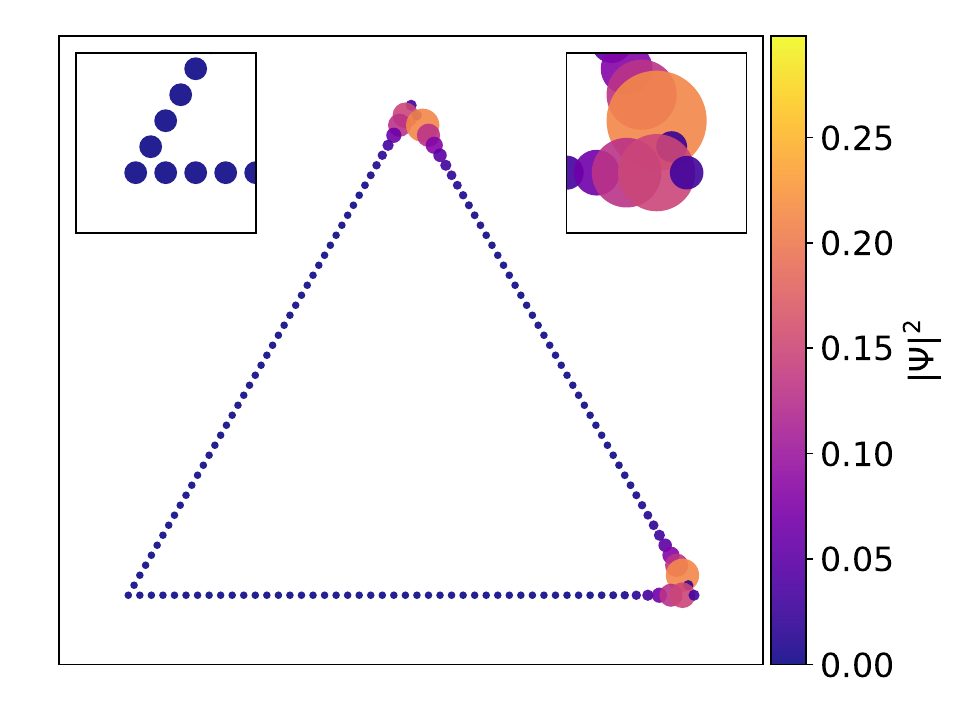}}\hspace{-30.0pt}
  \subfloat[]{\includegraphics[height=100pt]{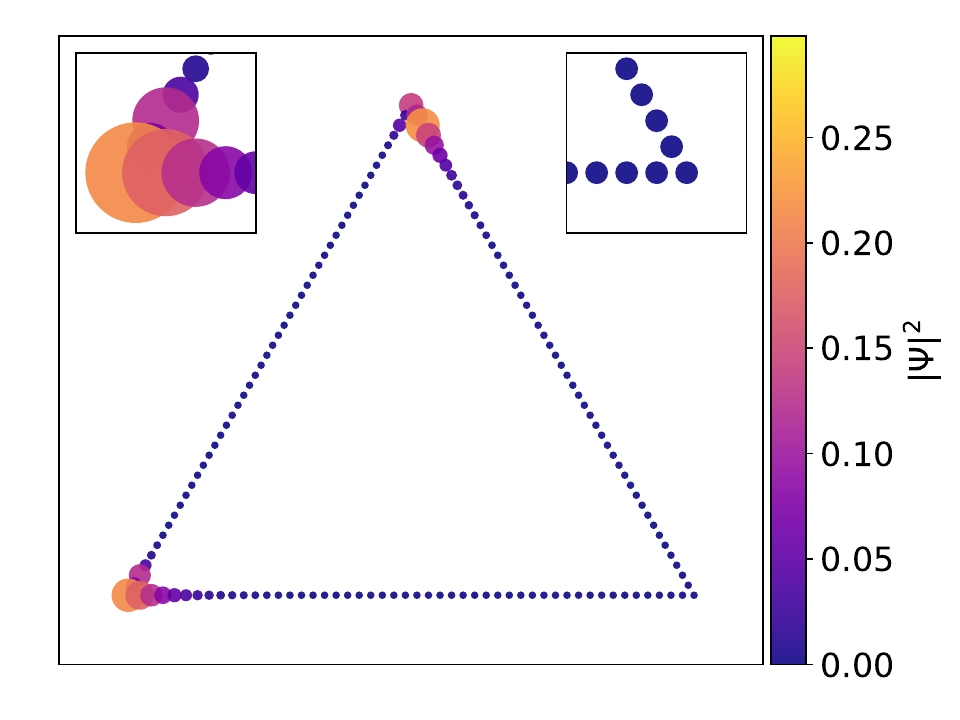}}
  \caption{(a) Topological phase diagram for a $W=1$ triangle by superimposing the $\mathcal{M}_{b,r,l}(A,\varphi)$ plots of 1D chains ($b$-bottom, $r$-right, $l$-left, $\mu=1.1$). $\varphi_{r,l}$ are equal to $\varphi_b+\pi/3$ and $\varphi_b-\pi/3$, respectively. The colors are coded by which edges have non-trivial topology. For example, Black---$[\mathcal{M}_b, \mathcal{M}_r, \mathcal{M}_l]=[1,1,1]$ means all edges are trivial. The behavior depicted in panels (b-f) is representative of that when $A$ is in the range of $(2.25,2.5)$, for which the $\mathcal{M}=-1$ phase ``crawls" through the three edges counterclockwise as $\varphi$ increases. (b) Spectral flow of a triangle with $W=1$, $L=50$, $\mu=1.1$, and $A=2.35$ with increasing $\varphi$. (c-f) BdG eigenfunction $|\Psi|^2$ summed over the two zero modes at $\varphi = 0$, $\frac{\pi}{12}$, $\frac{\pi}{6}$, and $\frac{\pi}{3}$, respectively. The bottom edge is parallel with $\hat{x}$ in the coordinates illustrated in Fig.~\ref{fig: pd} (a).}
  \label{fig: rotation}
\end{figure*}

In Fig.~\ref{fig: pd} (b) we show the topological phase diagrams for a 1D ribbon with width $W=1$, $\mathbf A = A\hat{y}$ and $\mathbf A = A(\frac{\sqrt{3}}{2}\hat{x}+\frac{1}{2}\hat{y})$ superimposed. We found that the vector potential component normal to the ribbon length direction has no effect on the Majorana number, nor does the sign of its component along the ribbon length direction. However, topological phase transitions can be induced by varying the size of the vector potential component along the ribbon, consistent with previous results \cite{romitoManipulatingMajoranaFermions2012, takasanSupercurrentinducedTopologicalPhase2022}. These properties motivate us to consider the structure of a hollow triangle formed by three finite-width ribbons subject to a uniform vector potential $\mathbf A = A\hat{y}$ as illustrated in Fig.~\ref{fig:triangles} (b), in which the bottom edge is aligned with $\hat{x}$. The purple regions on the phase diagram Fig.~\ref{fig: pd} (a) mean the bottom edge and the two upper edges of the hollow triangle have different $\mathcal{M}$, which should give rise to MZM localized at the two bottom corners if the triangle is large enough so that bulk-edge correspondence holds, and gap closing does not occur at other places along its edges.

To support the above arguments, we directly diagonalize the BdG Hamiltonian of a finite hollow triangle with edge length $L=50$ and width $W=1$. Fig.~\ref{fig: pd} (c) shows the spectral flow (BdG eigen-energies evolving with increasing vector potential $A$) close to zero energy at chemical potential $\mu=1.6$. Indeed, zero-energy modes appear in the regions of $\mu$ and $A$ consistent with the phase diagram. Hollow triangles with larger $W$ also have qualitatively similar behavior, although the phase diagrams are more complex \cite{supp}. The eigenfunctions for the zero-energy modes at $A=2.35$ and $\mu=1.1$ in Fig.~\ref{fig: rotation} (c) also confirm their spatial localization at the bottom corners of the triangle.

We next show that rotating the uniform vector potential in-plane, guided by the phase diagram of the three edges overlapped together [Fig.~\ref{fig: rotation} (a)], can manipulate the positions of the MZM. Specifically, a desired path on the $(A,\varphi)$ plane, $\varphi$ being the in-plane azimuthal angle of $\mathbf A$ [Fig.~\ref{fig: pd} (a)], of the phase diagram should make the nontrivial $\mathcal{M}=-1$ phase cycle through the three edges but without entering any trivial regions, when all edges have the same $\mathcal{M}$.

Fig.~\ref{fig: rotation} (b) plots the spectral flow versus $\varphi$ for a path determined in the above manner, which clearly shows that the zero-energy modes persist throughout the rotation and the bulk gap never closes. At a critical point when individual edges change their topology, e.g., near the middle of the $\varphi\in [0,\pi/6)$ region, gap closing is avoided due to finite-size effects, as discussed in \cite{aliceaNonAbelianStatisticsTopological2011}. Figs.~\ref{fig: rotation} (c-f) plot the BdG wavefunctions of the MZM at special values of $\varphi$. One can see that the two MZM appear to cycle through the three vertices by following the rotation of $\mathbf A$. We note in passing that if the vector potentials on the three edges can be controlled independently similar to the Kitaev triangle case, a swapping of the two MZM can in principle be achieved as well.

In \cite{supp} we also gave an example of a $W=3$ triangle, for which one has to additionally consider the nontrivial dependence of the bulk gap of the three edges on $\mathbf A$. In general, optimization of the parameter path can be done by examining the (suitably designed) topological phase diagram together with the bulk gap diagram, and choosing triangles of appropriate sizes.

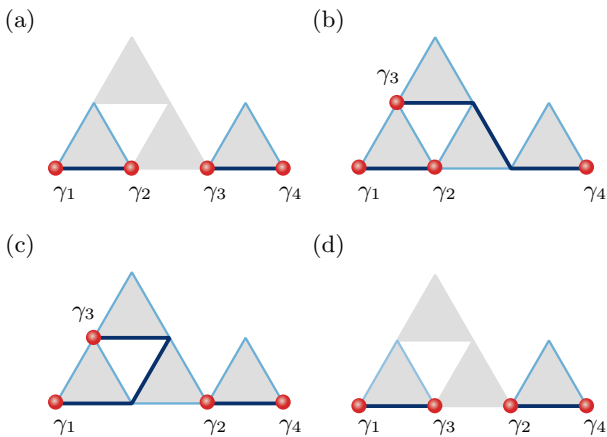
\begin{figure}[ht]
  \begin{tikzpicture}
    \node[inner sep=0pt] (figure) at (0,0)
    {\includegraphics[width=0.45\textwidth]{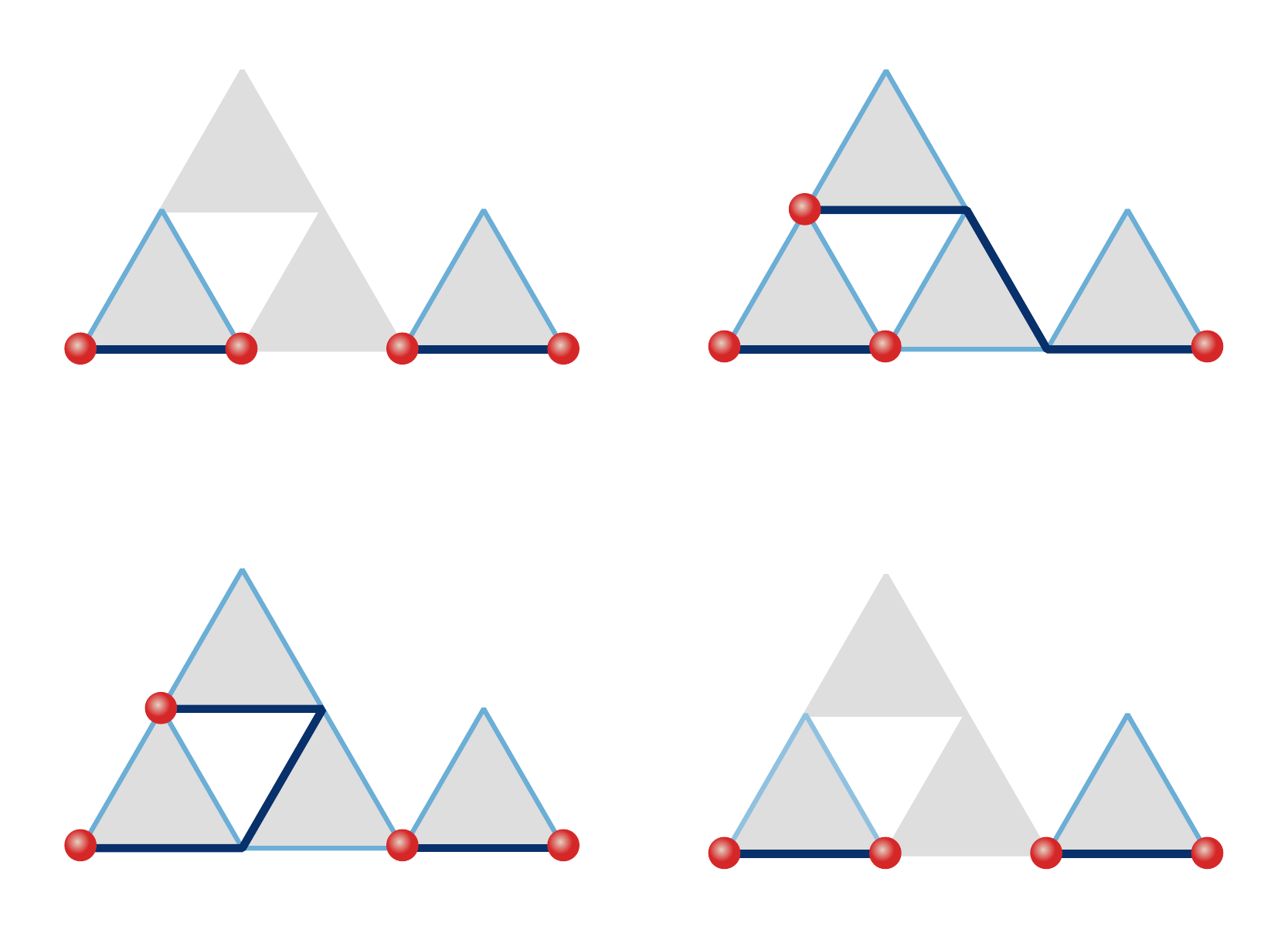}};

    \node[inner sep=0pt] (a) at (-4,2.75) {(a)};
    \node[inner sep=0pt] (b) at (0.1,2.75) {(b)};
    \node[inner sep=0pt] (c) at (-4,-.25) {(c)};
    \node[inner sep=0pt] (d) at (0.1,-.25) {(d)};

    \node[inner sep=0pt] (gamma1) at (-3.4,0.45) {$\gamma_1$};
    \node[inner sep=0pt] (gamma2) at (-2.4,0.45) {$\gamma_2$};
    \node[inner sep=0pt] (gamma3) at (-1.4,0.45) {$\gamma_3$};
    \node[inner sep=0pt] (gamma4) at (-0.4,0.45) {$\gamma_4$};

    \node[inner sep=0pt] (gamma1) at (0.65,0.45) {$\gamma_1$};
    \node[inner sep=0pt] (gamma2) at (1.65,0.45) {$\gamma_2$};
    \node[inner sep=0pt] (gamma3) at (0.9,2.00) {$\gamma_3$};
    \node[inner sep=0pt] (gamma4) at (3.65,0.45) {$\gamma_4$};

    \node[inner sep=0pt] (gamma1) at (-3.4,-2.65) {$\gamma_1$};
    \node[inner sep=0pt] (gamma3) at (-3.15,-1.15) {$\gamma_3$};
    \node[inner sep=0pt] (gamma2) at (-1.4,-2.65) {$\gamma_2$};
    \node[inner sep=0pt] (gamma4) at (-0.4,-2.65) {$\gamma_4$};

    \node[inner sep=0pt] (gamma1) at (0.65,-2.65) {$\gamma_1$};
    \node[inner sep=0pt] (gamma3) at (1.65,-2.65) {$\gamma_3$};
    \node[inner sep=0pt] (gamma2) at (2.65,-2.65) {$\gamma_2$};
    \node[inner sep=0pt] (gamma4) at (3.65,-2.65) {$\gamma_4$};

  \end{tikzpicture}
  \caption{Representative steps for braiding four MZM in four triangles sharing corners. (a) Initialization of four MZM $\gamma_1, \gamma_2, \gamma_3, \gamma_4$. All three edges of the bottom-middle and the top triangles are in the trivial phase by e.g. controlling the chemical potential. The bottom-left and bottom-right triangles have $\varphi = 0$ so that their bottom edges are nontrivial. (b) Moving $\gamma_3$ by ``switching on" the middle triangle by changing the chemical potential under a fixed vector potential at $\varphi=\frac{\pi}{6}$, and then turning on the top triangle with similar means except $\varphi = 0$. (c) Transporting $\gamma_2$ to the right triangle through rotating the vector potential in the middle triangle counterclockwise by $\pi/6$. (d) Moving $\gamma_3$ to the left triangle by ``switching off" the top triangle followed by the middle triangle.}
  \label{fig:4MZMbraiding}
\end{figure}

Before ending this section, we present a tentative design for braiding more than two MZM based on our hollow triangles. The structure, illustrated in Fig.~\ref{fig:4MZMbraiding}, consists of four triangles sharing corners with their neighbors. The critical step of transporting $\gamma_2$ to the left vertex of the rightmost triangle, corresponding to Figs.~\ref{fig:4MZMbraiding} (b,c), can be achieved by rotating the vector potential of the bottom-middle triangle counterclockwisely from $\varphi = \frac{\pi}{6}$ to $\frac{\pi}{3}$, which swaps the topological phases of the two side edges as shown in Fig.~\ref{fig: rotation}. In \cite{supp} we show this operation does not involve gap closing when the parameter path is chosen judiciously.

\emph{Discussion.---}The hollow interior of the triangles considered in this work is needed for two reasons: (1) $W\ll L$ is required for bulk-edge correspondence based on 1D topology to hold; (2) A finite $W$ is needed to gap out the chiral edge states of a 2D spinless $p$-wave superconductor. The latter is not essential if one does not start with a spinless $p$-wave supercondutor but a more realistic model such as the Rashba+Zeeman+$s$-wave pairing model. On the other hand, the former constraint may also be removed if one uses the Kitaev triangle. Nonetheless, an effective 3-site Kitaev triangle may emerge as the effective theory of triangular structures if a three-orbital low-energy Wannier basis can be isolated, similar to the continuum theory of moir\'{e} structures. We also note that the corner MZM in our triangles appear due to different reasons from that in higher-order topological superconductors \cite{wangEvidenceMajoranaBound2018,pahomiBraidingMajoranaCorner2020,zhangsb_2020_1,zhangsb_2020_2}.

For possible physical realizations of our triangles, immediate choices are quantum dots forming a Kitaev triangle \cite{dvirRealizationMinimalKitaev2023}, planar Josephson junctions or cuts on quantum anomalous Hall insulator/superconductor heterostructures \cite{xieCreatingLocalizedMajorana2021} that form a hollow triangle, and triangular atomic chains assembled by an STM tip \cite{schneiderPrecursorsMajoranaModes2022} on a close-packed surface. The quantum-dot platform may be advantageous in the convenience of implementing parity readout by turning the third vertex temporarily into a normal quantum dot \cite{mishmashDephasingLeakageDynamics2020,parity_QD_readout_2020, fengProbingRobustMajorana2022}. Looking into the future, it is more intriguing to utilize the spontaneously formed triangular islands in epitaxial growth \cite{pietzschSpinResolvedElectronicStructure2006} with the center region removed either physically by lithography/ablation, or electrically by gating. To create a staggered vector potential or supercurrent profile for the Kitaev triangle, one can use a uniform magnetic field, corresponding to a constant vector potential gradient, plus a uniform supercurrent that controls the position of the zero. It is also possible to use two parallel superconducting wires with counter-propagating supercurrents proximate to the triangle. Our work provides a versatile platform for manipulating MZM based on currently available candidate MZM systems and for potentially demonstrating the non-Abelian nature of MZM in near-term devices.

\begin{acknowledgements}
This work was supported by the start-up funding of CSU and partially by NSF CAREER grant DMR-1945023.
\end{acknowledgements}

\bibliography{triag_cite}

\end{document}